# Mixing of surface and bulk electronic states at a graphite-hexagonal boron nitride interface

Ciaran Mullan[1,†], Sergey Slizovskiy[1,2,†], Jun Yin[1,3,†], Ziwei Wang[1], Qian Yang[1,2], Shuigang Xu[2,4], Yaping Yang[1,2], Benjamin A. Piot[5], Sheng Hu[2,6], Takashi Taniguchi[7], Kenji Watanabe[7], Kostya S. Novoselov[1,2,8], A. K. Geim[1,2], Vladimir I. Fal'ko[1,2,9,*], Artem Mishchenko[1,2,*]

[1]*Department of Physics and Astronomy, University of Manchester, Manchester M13 9PL, UK.*
[2]*National Graphene Institute, University of Manchester, Manchester M13 9PL, UK.*
[3] *Nanjing University of Aeronautics and Astronautics, Nanjing 210016, China.*
[4]*Westlake University, Hangzhou 310024, China.*
[5]*Laboratoire National des Champs Magnétiques Intenses, LNCMI-CNRS-UGA-UPS-INSA-EMFL, 25 avenue des Martyrs, 38042 Grenoble, France.*
[6]*Xiamen University, Xiamen 361005, China.*
[7]*National Institute for Materials Science, 1-1 Namiki, Tsukuba, 305-0044, Japan.*
[8]*Centre for Advanced 2D Materials, National University of Singapore, 117546, Singapore.*
[9]*Henry Royce Institute for Advanced Materials, Manchester, M13 9PL, UK.*
[†]*These authors contributed equally*
*e-mail: vladimir.falko@manchester.ac.uk, artem.mishchenko@gmail.com

**Van der Waals assembly enables exquisite design of electronic states in two-dimensional (2D) materials, often by superimposing a long-wavelength periodic potential on a crystal lattice using moiré superlattices. Here we show that electronic states in three-dimensional (3D) crystals such as graphite can also be tuned by the superlattice potential arising at the interface with another crystal, namely, crystallographically aligned hexagonal boron nitride. Such alignment is found to result in a multitude of Lifshitz transitions and Brown-Zak oscillations for near-surface 2D states whereas, in high magnetic fields, fractal states of Hofstadter's butterfly extend deep into graphite's bulk. Our work shows a venue to control 3D spectra by using the approach of 2D twistronics.**

The bustling field of twistronics offers a level of tunability in two-dimensional (2D) van der Waals materials, otherwise unattainable in conventional systems [1-9]. One of the key concepts in twistronics is the twist-angle engineered moiré superlattice – long-wavelength periodic modulation of electronic potential resulting from the interference between two similar 2D lattices. Moiré superlattices spawn a plethora of novel physics including strong correlations and superconductivity in twisted bilayer graphene [10-12], resonant excitons, charge ordering, and Wigner crystallisation in transition metal chalcogenide moiré structures [13-18], Hofstadter butterfly fractal spectrum and Brown-Zak quantum oscillations in graphene crystallographically aligned with boron nitride [19-22], to name but a few. Besides 2D materials, twistronics can also be exploited for tailoring electronic states at the interfaces between multi-layered van der Waals crystals [23, 24].

At the surface of a crystal, the periodic lattice is interrupted. Surface states arise, as itinerant Bloch wavefunctions with real-valued crystal momenta, $k$, become evanescent states localised at the surface with complex $k_z$ values [25]. For example, surface charge accumulation in gapped semiconductors leads to distinct two-dimensional subbands easily tuned by electrostatic gating. While in metals, the high charge carrier density makes it difficult to observe and control surface states, as the bulk shunts the surface conductivity. Lying in between these two extremes are semimetals like bismuth and graphite, whose tunable surface states are interesting but not well explored yet. In this work, using van der Waals engineering and twistronics, we explore the gallery of moiré structuring of highly-tunable electronic states at the semimetal graphite surface, by aligning two bulk crystals, hexagonal graphite (Grt) and hexagonal boron nitride (hBN), to each other.

To this end, we prepared hBN/Grt/hBN heterostructures by aligning thin graphite films (typically, 5-10 nm thick) on top of hBN substrate and encapsulating the stack with another hBN crystal. Unless otherwise stated, this encapsulating hBN is intentionally misaligned (see Methods for details). As the lattice constants of hBN and graphite are similar, in a heterostack, they form a moiré superlattice with the periodicity controlled by the lattice mismatch, $\delta$ = 1.8%, and a misalignment angle, $\theta$ (Fig. 1A). In addition to providing moiré superlattice, the hBN encapsulation also preserves high electronic quality of graphite films [26-28]. Figure 1A-C shows schematics and micrographs of hBN/Grt/hBN heterostructures, fabricated into Hall bar and Corbino geometry devices (in total we have studied 10 graphite heterostructure devices). In these devices, top and bottom electrostatic gates were used to independently control carrier densities $n_t$ and $n_b$, at the top and bottom interfaces of hBN/Grt/hBN heterostructure.

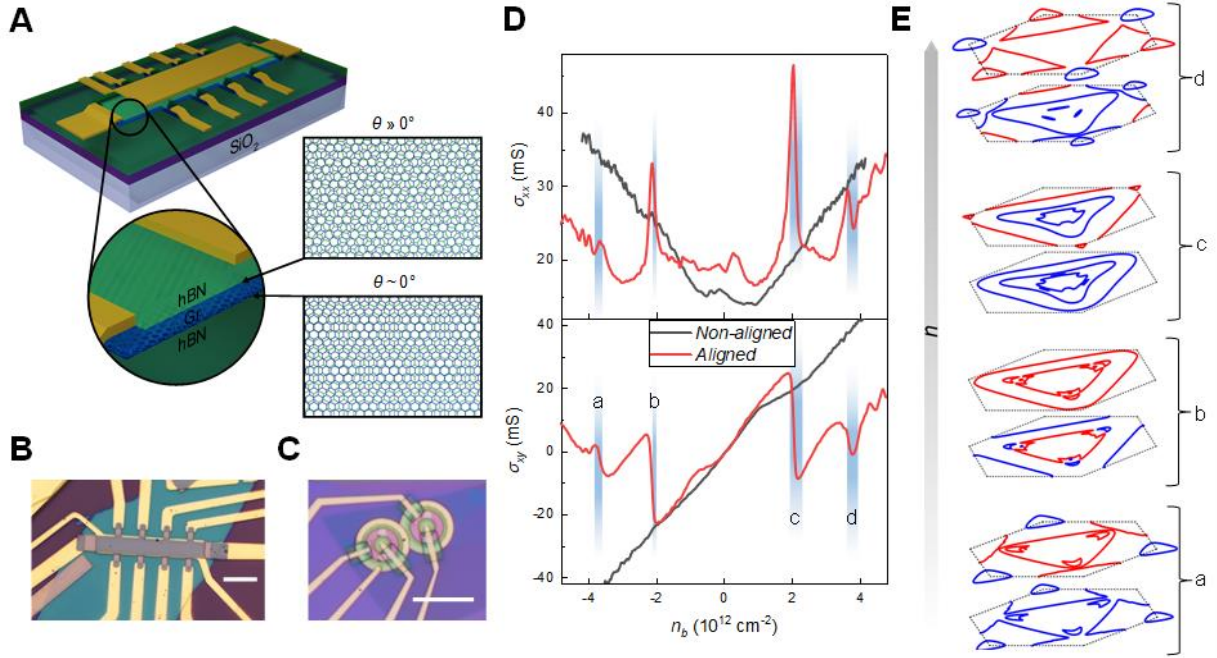

**Fig. 1. Moiré superlattice at the interface between graphite and hexagonal boron nitride.** (**A**) diagram of a typical heterostructure stack of graphite encapsulated in hBN with one of the interfaces aligned. Here the lattice mismatch between Grt and hBN has been exaggerated for clarity. (**B**, **C**) Optical micrographs of devices D1 (Hall bar geometry with 8.2-nm-thick graphite channel, bottom hBN aligned to graphite) and D3 (Corbino geometry with 5-nm-thick graphite channel, both top and bottom hBN flakes aligned to graphite). Scale bars are 10 μm. (**D**) Conductivities $\sigma_{xx}$ and $\sigma_{xy}$ as a function of carrier density induced by the bottom gate, $n_b$, for device D1 at the aligned interface and non-aligned device D4, measured at $T$ = 0.24 K and non-quantizing magnetic fields ($B$ = 120 mT). (**E**) Line cuts through the dispersion relation of the SBZ in the $k_x k_y$ plane, at carrier densities (bottom to top) $n$ = -3.8, -3.6, -2.1, -2.0, 1.9, 2.3, 3.6, 3.9 × $10^{12}$ cm$^{-2}$ grouped as pairs [labels *a,b,c,d* correspond to the regions highlighted in (D)]. Black dashed hexagon denotes the boundary of the 1$^{st}$ SBZ and red (blue) denotes holes (electron) line cuts, some lines at the corners are extended into 2$^{nd}$ SBZ for clarity.

Hexagonal graphite (Bernal stacking) is a compensated semimetal with Fermi surface occupying only a small fraction of Brillouin zone. The size of the Fermi surface is determined mostly by a through-layer hopping parameter, $\gamma_2$ ≈ -20 meV [29]. Due to its semimetallic nature, graphite does not host surface states (evanescent modes) in the absence of dangling bonds or electric fields on its surfaces. However, when electric field above a critical value is applied perpendicular to graphite basal plane, tunable surface states start to appear, see refs. [26, 30] and Supplementary Note 1.

Moiré superlattice drastically modifies graphite surface states, resulting in entirely different transport behaviour between aligned and non-aligned devices, Fig. 1D. Non-aligned graphite interface shows nearly featureless carrier density dependence of longitudinal $\sigma_{xx}(n)$ and transversal $\sigma_{xy}(n)$ conductivities in small magnetic fields. However, when gating aligned graphite interface, $\sigma_{xy}(n)$ displays multiple zero crossings accompanied by the corresponding peaks in $\sigma_{xx}(n)$. We attribute these crossings to the recurrence of electrostatically induced surface states dominated by electron- or hole-like charge carriers. To quantify this, we calculated Fermi surface projections using an effective-mass model with Slonczewski-Weiss-McClure (SWMC) parameterisation of graphite [31, 32] subjected to moiré superlattice potential, in combination with self-consistent Hartree analysis (see Methods). Our calculations in superlattice Brillouin zone (SBZ) of hBN/Grt/hBN heterostructure show kaleidoscopic surface states with numerous topological Lifshitz transitions (LTs) across a range of carrier densities, shown in Fig. 1E. Pairs of plots (Fig. 1E, a-d) with drastic changes in the Fermi surface topology demonstrate four LTs, corresponding to the four carrier density ranges highlighted in Fig. 1D. LTs observed at $|n| \approx 2$ and at $|n| \approx 3.7 \times 10^{12}$ cm$^{-2}$ belong to two different branches of surface states, the former ones mostly residing on the first graphene bilayer in Grt, while the latter are more peaked at the second bilayer (Fig. S1C). As magnetic field increases the surface states in the vicinity of LTs give rise to separate branches of Landau levels; the evolution of $\sigma_{xx}(n)$ and $\sigma_{xy}(n)$ in low magnetic fields is further illustrated in Supplementary Note 2.

At higher $B$ fields, the difference between aligned and non-aligned interfaces of hBN/Grt/hBN system becomes even more prominent. Figure 2A plots $\sigma_{xx}$ as a function $B$ for surface states on aligned and non-aligned interfaces. The curves were measured at $T = 60$ K to supress the effects of Landau quantisation. When the aligned surface is doped away from electron-hole compensation point, $\sigma_{xx}$ exhibits oscillatory behaviour periodic with $1/B$. Peaks in $\sigma_{xx}$ can be traced to values of $B_{1/q} = \frac{1}{q}\frac{\phi_0}{A_0}$ corresponding to integer number $q$ of superlattice unit cells $A_0 = \sqrt{3}/2\,\lambda^2$, commensurate with a magnetic flux quantum $\phi_0 = h/e$, here $\lambda$ is the wavelength of moiré superlattice, $h$ is the Planck's constant, and $e$ is the elementary charge. The commensurability between $\phi_0$ and magnetic flux through a moiré unit cell $\phi = BA_0$ can be interpreted as the manifestation of Brown-Zak quantum oscillations at the superlattice interface, which were recently reported for aligned graphene/hBN heterostructures [21, 22]. Similarly, the formation of magnetic Bloch states leads to higher conductivity due to straight rather than cyclotron trajectories of surface charge carriers [21, 22, 33, 34], as evidenced by the conductivity peaks at $B_{1/q}$, Fig. 2A. Figure 2B shows these $n_b$-independent conductivity peaks located at all distinguishable $1/q$ commensurate fields. Note that not only unit fractions but also second-order fractal states (e.g., $B_{2/5}$ in Fig. S3A) are evident in Brown-Zak oscillations.

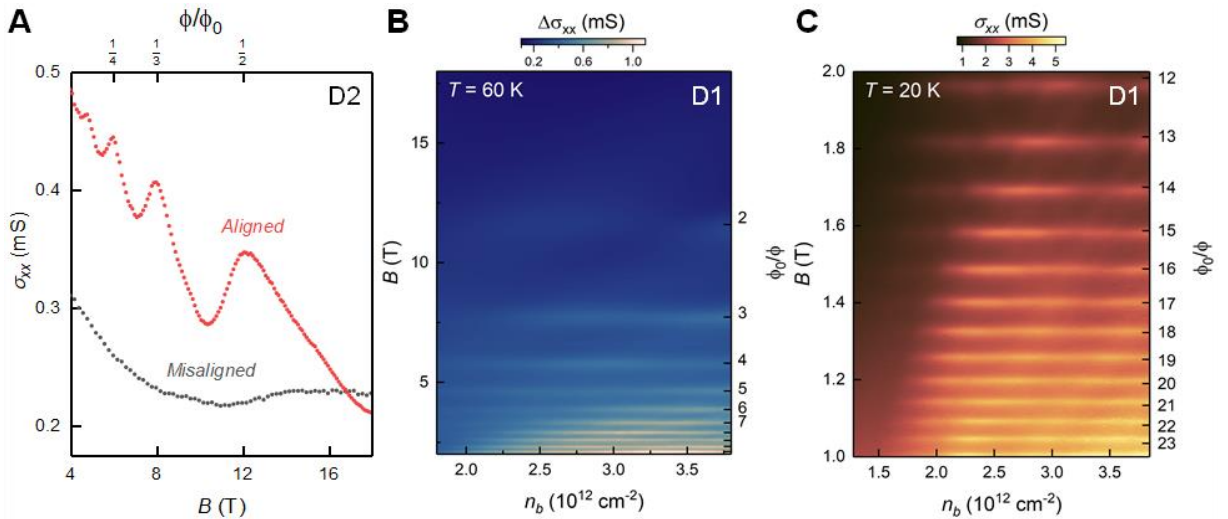

**Fig. 2. Brown-Zak oscillations arising from states confined to the graphite-hBN interface.** (**A**) Conductivity $\sigma_{xx}$ as a function of $B$ for device D2 (Corbino geometry with 7-nm-thick graphite channel) at high electron doping electrostatically induced by top (bottom) gate, which tunes the aligned (non-aligned) graphite-hBN interface. When tuning the aligned interface, peaks are evident at values of $B$ equivalent to flux quantum $1/q$ per superlattice unit cell. Carrier density $n_t = 3.1 \times 10^{12}$ cm$^{-2}$ and $n_b = 3.3 \times 10^{12}$ cm$^{-2}$ for aligned and non-aligned interfaces, respectively. (**B**) Conductivity map (a smooth background subtracted, see Methods) as a function

of $B$ and $n_b$, for D1 at the aligned graphite-hBN interface. Measurements were performed at $T$ = 60 K to suppress the Landau quantisation, though some features corresponding to surface Landau bands still persist (see Fig. S3 for full range mapping). Right hand axis denotes the inverse flux $\phi_0/\phi$. (**C**) $\sigma_{xx}(n_b,B)$ map on the same device at intermediate temperature, $T$ = 20 K.

Since Brown-Zak oscillations stem from translational invariance of magnetic Bloch states at rational fractions of magnetic flux $\phi/\phi_0 = p/q$, they are expected to be insensitive to temperature as long as electrons retain phase coherence within magnetic supercell $qA_0$. Figure 2C shows that at intermediate temperatures (T = 20K), states with $\phi_0/\phi$ = 24 are clearly visible (and $\phi_0/\phi$ > 35 states are distinguishable in Fig. S3), providing a lower bound on phase coherence length > 85 nm. Brown-Zak oscillations can also be interpreted as the Aharonov-Bohm interference of electron trajectories in a periodic 2D network formed by classical trajectories of electron drifting around the Fermi contours in a magnetic field that are joined by magnetic breakdown tunnelling in the vicinity of Van Hove singularities [35]. This interpretation allows a convenient transition to the regime of low $B$-fields where we see multiple Lifshitz transitions of the Fermi surface topology, as shown in Fig. 1E, and explains the disappearance of Brown-Zak oscillations for $|n_b|$< 2 × $10^{12}$ cm$^{-2}$.

In comparison, gating the non-aligned interface in our hBN/Grt/hBN heterostructures does not result in Lifshitz transitions or Brown-Zak oscillations (cf. Fig. 1D, Fig. 2A, and Supplementary Note 3). This is not surprising, as our previous study has shown that the surface states on the two surfaces of a graphite film are well screened from each other, with the screening depth of the order of 2-3 layers [26]. Raman measurements also do not show any qualitative difference in strain distribution or effects of alignment for films thicker than 7-8 graphene layers at the aligned and non-aligned graphite surfaces (see Supplementary Note 4). Our results are further supported by a recent paper on atomic relaxation in multilayer moiré heterostructures, which predicts very short (1 layer) penetration depth of moiré reconstruction for superlattices with $\lambda$ < 20 nm [23].

However, to our surprise, when we cooled our devices to base temperature (30 mK) and measured Landau fan map, we observed a development of Hofstadter butterfly – the fractal quantum Hall effect (QHE), not just in the surface states, but across the entire bulk of the graphite film, Fig. 3. A low-$T$ high-$B$ Landau fan map of conductivity $\sigma_{xx}$ versus $n = n_t + n_b$ for device D2, presented in Fig. 3A, reveals multiple QHE features. Figure 3C summarises the results in a Wannier diagram. Analogous map and Wannier diagram for device D3 are plotted in Fig. 3B, D. Although QHE is in principle forbidden in a 3D electronic system like graphite, it has been recently reported for thin (up to 100 nm) graphite films [26]. Two main factors contribute to the observed QHE: dimensional reduction of electronic system from 3D semimetal to 1D Landau bands in strong $B$-fields, and the consequent formation of standing waves in the 1D Landau bands due to finite thickness of graphite film. Standing waves result in the quantisation of 1D Landau bands and the development of minigaps among them, manifesting as a special 2.5D QHE. At large $B$-fields (above those marked by white dashed lines in Fig. 3A, B), only the two lowest Landau bands (0 and 1) remain crossing the Fermi level and contribute to the transport. These two bands are split by magnetic field $\delta_{10}$ ≈ 0.4 meV/T. In addition to being gapped by the standing waves, they are also spin resolved by Zeeman effect, $2\mu_B B$ ($\mu_B$ is the Bohr magneton), while lifting their +KH and −KH valley degeneracy depends on the graphite layer parity.

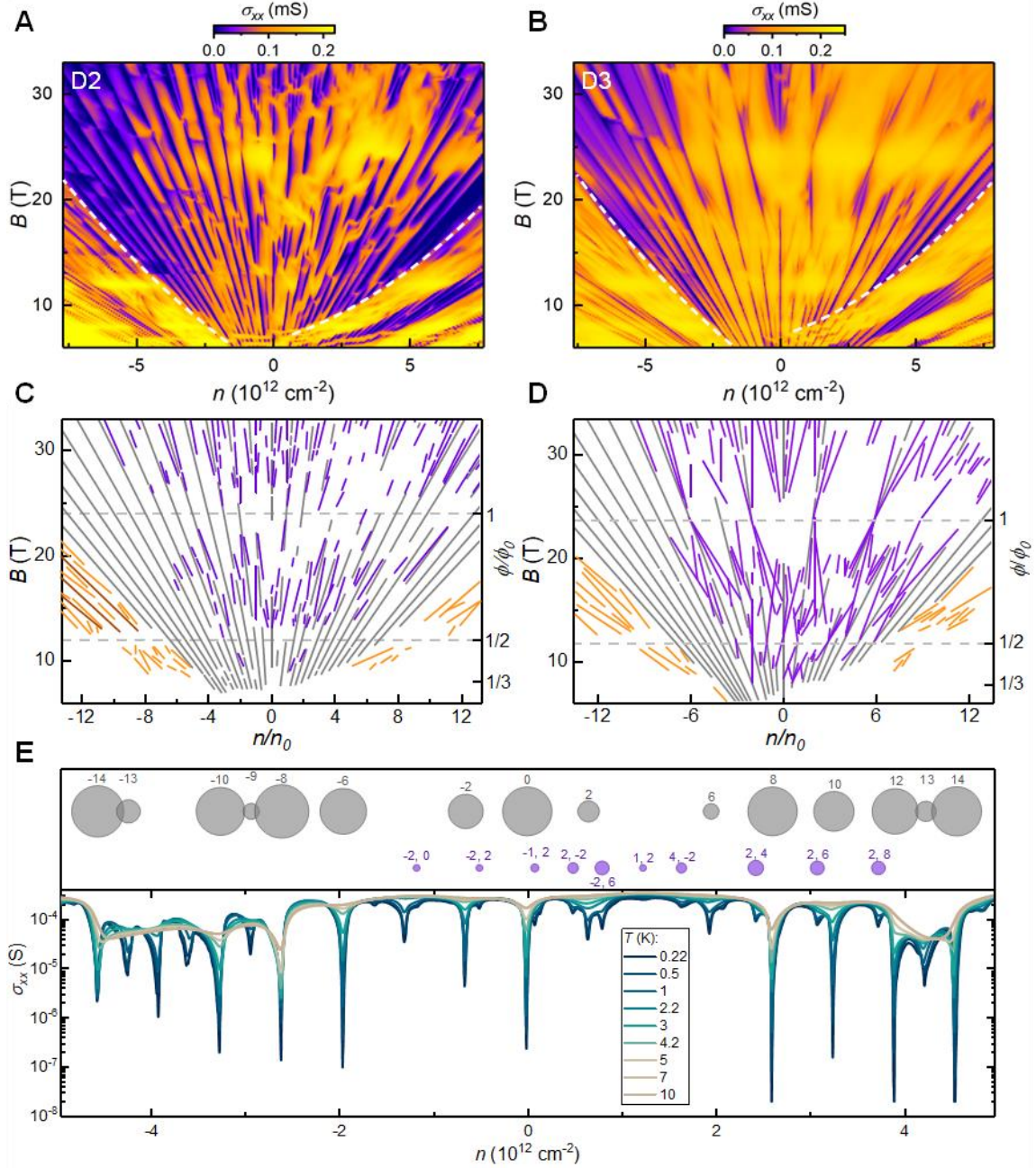

**Fig. 3. Fractal QHE states in graphite.** (**A, B**) Conductivity $\sigma_{xx}$ as a function of $n = n_t + n_b$ and $B$ for devices D2 and D3 measured at $T$ = 30 mK and $n_t = n_b$ (i.e., displacement field $D$ = 0). The white dashed lines denote transition from surface Landau levels to bulk ultraquantum regime (UQR). (**C, D**) Associated Wannier diagrams plotting the QHE (grey) and fractal QHE (purple) states in the UQR observed in (A) and (B). Horizontal axis is normalised in units of electron per supercell, with $n_0 = 1/A_0$. Below UQR, orange (brown) lines trace fractal (non-fractal) states in surface Landau levels +2 and -2. (**E**) Hierarchy of QHE gaps in aligned hBN/graphite/hBN heterostructure. Lower panel shows $\sigma_{xx}(n)$ traces at different temperatures for device D3 at $B$ = 13.5 T, which is used to extract gap sizes from Arrhenius activation. Upper panel is a bubble plot of QHE gaps; the area of a circle scales linearly with the gap size (ranging from 30 μeV to 1.8 meV). Colour coding is the same as in (D) and labels are integers $s$ and $t$ from Eq. 1, for the QHE ($t$ only) and fractal QHE ($s,t$) states.

The Hofstadter butterfly [36] – a fractal set of energy eigenvalues for magnetic flux $\phi/\phi_0 = p/q$ commensurate with moiré periodic potential – is plotted in Fig. 4A for a honeycomb lattice [37], matching the geometry of our moiré perturbation. In charge transport measurements, this fractal set manifests in the Diophantine equation for the Landau fan and its Wannier diagram representation:

$$\frac{n}{n_0} = t\frac{\phi}{\phi_0} + s \qquad (1)$$

Here integer *t* is the Landau filling index $v = nh/eB$ and integer *s* is the superlattice Bloch band filling index; $n_0 = 1/A_0$ is the charge carrier density per superlattice unit cell area. For *s* = 0, Eq. 1 plots conventional Landau fan diagram with $t \equiv v$ (grey lines in Fig. 3C, D), while for $s,t \neq 0$, it traces Hofstadter states (purple lines in Fig. 3C, D) emanating from *B*-fields satisfying $\phi/\phi_0 = p/q$.

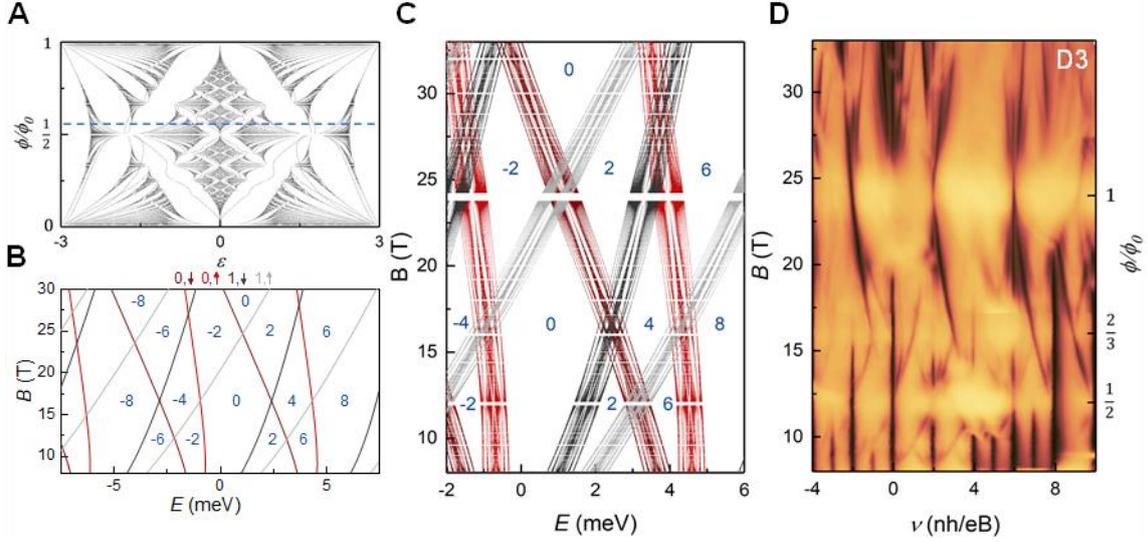

**Fig. 4. Hofstadter broadening of energy levels in graphite.** (**A**) Hofstadter butterfly calculated for a honeycomb lattice following Ref. [37], where for simplicity the hopping parameter *γ* = 1. The dashed line denotes a flux ratio of $\phi/\phi_0$ equivalent to *B* = 13.5 T in device D3, which is the field strength as in Fig. 3E. (**B**) Allowed energy levels resulting from quantised states from 0 (1) Landau bands shown in red (grey), calculated for 16-layer-thick graphite film without a moiré perturbation [26]. Zeeman splitting has been included, as indicated by light and dark lines for spin up and spin down, respectively. Integer labels in blue refer to filling factor *v*. (**C**) Combination of panels (A) and (B) by applying the Hofstadter butterfly as a small perturbation to each energy level, Eq. 2. Labelled *v* same as in (B). (**D**) Conductivity map for device D3, similar to that in Fig. 3B except plotted as a function of filling factor of main QHE sequence. Colour scale brown to yellow is from 0 to 247 μS.

Figure 3E shows a clear hierarchy of QHE gaps: those measured at integer filling factors (*s* = 0) are an order of magnitude larger than the gaps between Hofstadter butterfly states ($s,t \neq 0$). This suggests that the effect of moiré superlattice on the QHE can be considered as a small perturbation. To estimate the energy scale of this perturbation, we envelop 0 and 1 Landau bands in graphite with a Hofstadter butterfly energy spectrum. The spectrum of Landau levels for a 16-layer-thick graphite without moiré perturbation calculated following [26] (also see Methods) is plotted in Fig. 4B. It shows how, with increasing *B* field, levels cross each other, corresponding to closure and subsequent reopening of QHE gaps. Figure 4C plots the energy levels with the moiré envelope incorporated by giving each Landau level $E_m$ of the graphite QHE the fine structure of the Hofstadter butterfly *ε*:

$$E_m^{\text{moiré}} = E_m + S\varepsilon \qquad (2)$$

Here $S \approx 0.14$ meV is the scaling factor for the bandwidth of Hofstadter butterfly [38], which was estimated from the measured transport gaps. The obtained enveloped spectrum shows good agreement with the experimental data as a function of filling factor, in terms of gap sizes and positions, as plotted in Fig. 4D.

We have shown that surface states of graphite (and, potentially, other semimetals) can be efficiently tuned by moiré superlattice potential resulting in Brown-Zak oscillations and Hofstadter surface states, stemming from aligned graphite and hBN interface. The alignment provides kaleidoscopic Lifshitz transitions, which develop into Brown-Zak oscillations and Hofstadter surface states. Remarkably, at low temperatures and high magnetic fields, the periodic surface potential leads to a bulk fractal Hofstadter butterfly which penetrates the entire graphite film. Hofstadter spectrum in this bulk twistronics system offers enticing possibility to explore bulk-boundary correspondence in the QHE regime.

## Methods

### Device fabrication

To make Hall bar devices, graphite flakes were encapsulated by hBN through dry transfer described elsewhere [39, 40]. Briefly, hBN was first mechanically exfoliated onto $SiO_2$/Si substrates. The target hBN flake was then picked up by polymethylmethacrylate (PMMA) film and released onto a graphite flake exfoliated on $SiO_2$/Si substrate with a partially overlapped area. To make aligned hBN/graphite structures, the straight edges of hBN and graphite flakes, which are usually along their crystallographic axes, are aligned in parallel. Standard e-beam lithography, reactive ion etching, and e-beam evaporation were applied sequentially to the hBN/graphite/hBN stack to pattern top gate electrode and metal contacts to graphite flake (3 nm Cr/80 nm Au). The device was then shaped into Hall bar geometry using a thermally evaporated aluminium film as etch mask, which was later removed by 0.1 M NaOH solution. To make devices in Corbino geometry, we overexposed the resist to form crosslinked PMMA bridge, to separate inner contact, top gate, and the outer contact.

Graphite capacitors used to study surface states of non-aligned heterostructures, were fabricated similarly, but with quartz substrate replacing $SiO_2$/Si wafer to minimise the parasitic capacitance. For capacitance devices, we used graphite flakes around 50 nm thick, guaranteeing the 3D graphite electronic spectrum. Relatively thick hBN flakes (> 40 nm) were chosen to eliminate the inhomogeneity of electrostatic potential introduced by relatively rough metal electrode.

### Transport and capacitance measurements

The longitudinal and Hall voltages of the Hall bar device were recorded with lock-in amplifiers (SR830 or MFLI) while applying a small low-frequency ac current of 10 nA (except where higher current is specified). For Corbino devices, a small ac bias (40-100 µV) was applied to the inner contact, while current was recorded from the outer one.

Differential capacitance $C$ was measured as a function of bias voltage $V_b$ between the metal gate and graphite using an on-chip cryogenic bridge [41], which reaches sensitivity ≈10 aF at 1 mV excitation. In general, ac excitations of 102.53 kHz frequency and opposite phases were applied to the sample and a reference capacitor of known capacitance, respectively. Output signals from these two capacitors were mixed and applied to the gate of a high electron mobility transistor (HEMT), which served as an amplifier. The excitation voltage of the reference capacitor was modulated so that the output signal from the HEMT became zero, and the capacitance of the sample can be deduced based on the ratio of their excitation voltage at the balanced point. The typical excitation voltage applied to the samples ranged from 1 to 10 mV, depending on the thickness of the hBN dielectric layer.

### Tight-binding description of surface states

To describe surface states, we focus of the evanescent modes with complex momentum $k_z$. To this end, we solve analytically the spectrum of graphite [32], considering the boundary conditions ($\psi$ = 0 for surface carbon atoms). There are no complex $k_z$ solutions at zero doping, as only real $k_z$ solutions satisfying zero boundary conditions are normalizable. Electrostatic doping of graphite surface creates an inhomogeneous in z-direction potential near the surface, which does not preserve $k_z$ momentum, allowing real $k_z$ solutions near the surface, which then turn into evanescent modes decaying into the bulk. This provides a heuristic picture for the origin of non-trivial surface state solutions, Fig. S1D,E.

To quantify the surface states, we numerically find the self-consistent potential profile of doped graphite surface, considering a finite-thickness graphite sandwiched between two gates with carrier densities $n_t$ and $n_b$. Since graphite screens the electric field at a depth of few layers, it is equivalent to a semi-infinite bulk graphite with one surface for $N \geq 10$, where $2N$ is the number of graphene layers. In the Hartree approximation, the potentials on the layers, $U_j > 1$, are related with layer electronic densities as

$$U_j = U_1 + \frac{e^2 c}{\epsilon \epsilon_0}\left[(j-1)n_b + \left(\sum_{p=1}^{j-1}\sum_{l=p+1}^{2N} n_l\right) - \frac{\epsilon-1}{4}(n_1 - n_j)\right], \quad (3)$$

Where $\epsilon$ = 2.6 accounts for vertical polarizability of graphene [42], $c$ = 3.35 Å is interlayer separation.

We temporarily fix the value of $U_1$, which plays the role of a surface chemical potential, and then self-consistently calculate Hartree potentials and densities on all the layers. The electronic density in layer $l$ of graphite, calculated in the Hartree approximation, is

$$n_l = 2\int_{BZ} \frac{d^2\vec{k}}{(2\pi)^2} \sum_{n=1}^{4N} f\left(\epsilon_n(\vec{k})\right)\left(\left|\Psi_n^A(l,\vec{k})\right|^2 + \left|\Psi_n^B(l,\vec{k})\right|^2\right) - n_0, \quad (4)$$

where $f$ is a Fermi-Dirac distribution, and $n$ enumerates the eigenfunctions for a given in-plane momentum $\vec{k}$. The constant $n_0$ is chosen to match $\sum_{l=1}^{2N} n_l = 0$ to provide electrical neutrality. After finding the densities on all the layers we relate $U_1$ with $n_t$ via $n_t = -n_b - \sum_{n=1}^{2N} n_l$. Numerically stable self-consistent solutions of Eq. 3 and 4 for different values of $U_1$ can also be used to calculate the quantum capacitance through Eq. 5 in Supplementary Note 1.

*Tight-binding description of surface states in aligned graphite*

The spectrum for graphite aligned with hBN was calculated by treating the periodic moiré potential as a perturbation applied only to the top graphene layer, we followed the standard procedure [43], using the mirror-symmetric superlattice coupling Hamiltonian

$$\delta H = \sum_{m=0}^{5} e^{ig_m \cdot r}\left[U_0 + (-1)^m\left(i\, U_3\sigma_3 + U_1 \frac{g_m \times \hat{z}}{|g|}\cdot \sigma\right)\tau_3\right]$$

applied to the two top-layer components of the graphite film wavefunction, where Pauli matrices $\sigma$ operate on top layer sublattices and $\tau$ operates on valleys, $g_m = \mathcal{R}_{\pi(m-1)/3}\{0, 4\pi\delta/(3a)\}$ are 6 reciprocal lattice vectors of superlattice (where $\mathcal{R}$ is a rotation matrix) $\delta = 0.018$ is a lattice mismatch $a = 1.42$ Å is C-C distance and we use the parameters $U_0$ = 8.5 meV, $U_1$ = -17 meV, $U_3$ = -14.7 meV [44, 45]. The results do not significantly depend on the values of superlattice couplings, and it turned out to be sufficient to restrict the momentum space to the first star of superlattice reciprocal lattice vectors to achieve convergence.

*Calculation of $\Delta\sigma_{xx}$ in high temperature mappings*

To highlight Brown-Zak oscillations across a large range of magnetic fields, we calculated $\Delta\sigma_{xx}$ by subtracting a smooth background from $\sigma_{xx}$ data. In comparison to graphene-hBN systems where the background conductivity can be fitted with polynomials [21], we find even high order (>10) polynomials insufficient as many oscillatory artefacts are present. Instead, we employ a two carrier Drude model of $\sigma_{xx}(B)$, $\sigma_{xy}(B)$ and fit both simultaneously to yield carrier densities and mobilities $n_h$ = 2.2 ×10$^{12}$ cm$^{-2}$, $\mu_h$ = 24,000 cm$^2$V$^{-1}$s$^{-1}$, $n_e$ = 2.8 ×10$^{12}$ cm$^{-2}$, $\mu_e$ = −19,000 cm$^2$V$^{-1}$s$^{-1}$ for zero gate bias at $T$ = 60 K. This two carrier model fit, $\sigma_{xx}^{fit}(B)$, is then used to calculate $\Delta\sigma_{xx}(n_b,B) = \sigma_{xx}(n_b,B) - \sigma_{xx}^{fit}(B)$. Oscillations observed in $\Delta\sigma_{xx}$ were cross examined against $\sigma_{xx}$ to confirm they are not artefacts.

*Tight-binding description and standing waves in graphite in UQR*

The allowed energy levels of graphite bulk in the absence of moiré perturbation were calculated following Ref. [26], and using the SWMC parameterisation of [31, 32] to generate the zero-field spectrum of graphite. The $B_z$-field leads to Landau quantization of the electron motion in plane and forming Landau bands. The UQR is defined as the regime where only the two lowest Landau bands (0 and 1) cross the Fermi level at $k_F$ ≈ $\pi/4c$, with interlayer separation $c$ = 0.335 nm. In thin graphite films these 0 and 1 Landau bands are further discretised by the formation of standing waves in the $k_z$ direction, where the number of states these 1D bands split into (and, hence, the energy spacing between these states) is determined by the number of layers $N$. Further to this, Zeeman splitting, $2\mu_B B$, is applied to the resultant states, in order to complete the plot for 16-layer graphite in Fig. 4B.

*Analysis of Hofstadter spectrum and Brown-Zak oscillations*

To model the conductivity of our aligned devices at high *B* field and low temperatures we consider the Moiré superlattice potential as a weak perturbation; each Landau-level-like energy level described above (and plotted in Fig. 4B) is split into *q* subbands, at a given $\phi/\phi_0 = p/q$. At rational fields $\phi/\phi_0 = p/q$ translational symmetry is restored if we consider a magnetic unit cell of *q* supercells, and the *q*×*q* tight binding Hamiltonian is defined by the difference equation [37]

$$(\varepsilon^2 - 3)\psi_m = 2\cos\left[\pi\frac{\phi}{\phi_0}\left(m+\frac{1}{2}\right)+\kappa\right]\psi_{m+1} + 2\cos\left[2\pi\frac{\phi}{\phi_0}\left(m+\frac{1}{2}\right)+2\kappa\right]\psi_m \\ + 2\cos\left[\pi\frac{\phi}{\phi_0}\left(m-\frac{1}{2}\right)+\kappa\right]\psi_{m-1}$$

where *m* is an integer, $\kappa = k_y a\sqrt{3}/2$, and the calculation has been simplified by using a normalised hopping parameter *γ* = 1, which yields an energy range of *ε* = ±3 (arb. units). Solving for energy eigenvalues generates the honeycomb lattice Hofstadter butterfly (Fig. 4A). Note that near significant flux fractions $\phi/\phi_0$ = 1, 1/2, 1/3 the density of points increases, and much larger *q* are required to populate these regions, so to keep computation time reasonable the calculations are limited to *q* ≤ 100.

In order to apply this fine structure of Hofstadter butterfly to each energy level we need a relevant energy scaling factor, defined as *S* in the main text. Analysis of the thermal activation of gaps for device D3 at *B* = 13.5 T (Fig. 3E) indicate the largest fractal gaps are ≈0.1 meV (consistent between Corbino devices, see Supplementary Note 3 and Fig. S4). We then attribute this to the largest gap in the Hofstadter spectrum at the flux value $\phi/\phi_0$ = 0.56 (corresponding to *B* = 13.5 T, dashed line in Fig. 4A), which occurs at 1.0 < *ε* < 1.7 (or -1.7 < *ε* < -1.0 due to Hofstadter butterfly symmetry) therefore resulting in *S* = 0.14 meV. The full spectrum is then calculated using Eq. 2 and plotted in Fig. 4C where we use the periodicity of Hofstadter butterfly (such that $\varepsilon(\phi/\phi_0 + \rho) = \varepsilon(\phi/\phi_0)$ for any integer *ρ*) to plot states at $\phi/\phi_0$ > 1. The result is not only the presence of fractal fine structure in the spectrum but also the reduction in size of the main QHE gaps, due to Hofstadter butterfly effectively broadening each energy level by a finite width of 6*S*. Again, the computation limit of *q* ≤ 100 results in a blank region at $\phi/\phi_0$ = 1, which are in fact the densest regions. The overlap of these dense regions at $\phi/\phi_0$ = 1 completely closes some of the main QHE gaps (e.g., *ν* = 0 and 4) which is in good agreement with the measured conductivity (Fig. 4D).

# Supplementary information

Supplementary Note 1 – Surface states in non-aligned graphitic films in zero *B*-field

To better understand the origin of the surface states in non-aligned graphite, we performed capacitance measurements in devices comprised of graphite covered with non-aligned hBN of a thickness *d*, and a metal gate, as shown in the inset of Fig. S1A. Capacitance spectroscopy is a powerful tool to probe thermodynamic density of states (*DoS*) at the interfaces, and it has been successfully applied to study two dimensional systems [41, 46]. However, its application to study surface states of metals or semimetals is rare. The measured capacitance (*C*) can be considered as geometric parallel-plate capacitance $C_G = \varepsilon\varepsilon_0 A/d$ and quantum capacitance $C_Q$ in series, $1/C = 1/C_G + 1/C_Q$ [47]. The quantum capacitance reflects the *DoS* = $dn/dU_S$ on the surface of graphite: $C_Q = Ae^2 dn/dU_s$, where *A* is device area, *e* is the electron charge, *n* is the carrier density, and $U_s$ is the surface chemical potential (labelled $U_1$ in the Methods section). In our measurements, *C* roughly follows a V-shaped dependence on *n*, where $n(V_b) = \frac{1}{Ae}\int_0^{V_b} C(V)dV$, with a notable fine structure, Fig. S1A.

To analyse our data, we adapted an effective-mass model of 14-layer-thick graphite film with the SWMC parametrisation of graphite [31, 32] combined with self-consistent Hartree potentials to account for electrostatic gating, for details see Methods and Ref. [26]. From the self-consistently calculated Hartree potentials and carrier density on each layer, the capacitance (per unit area) can be written as

$$C^{-1}(n) = C_G^{-1} + \left(-e\frac{dn}{dU_s}\right)^{-1}\bigg|_{U_s=U_s(n)} \qquad (5)$$

By comparing the calculated capacitance with experimental data, we obtained a set of SWMC parameters. Results of such procedure are shown in Fig. S1B, where experimental data from four different devices are compared to a fitted theoretical *DoS*($U_s$) = $dn/dU_s$ curve, showing excellent agreement between theory and experiment.

The dispersion spectrum for a 14-layer graphite film with self-consistently determined layer potentials is shown in Fig. S1C for $n = 6\cdot 10^{12}$ cm$^{-2}$. Red colour in Fig. S1C indicates higher amplitude of the corresponding state at the surface graphene bilayer. For instance, the two red-coloured branches seen in Fig. S1C would be absent in charge-neutral graphite, where no surface states are present.

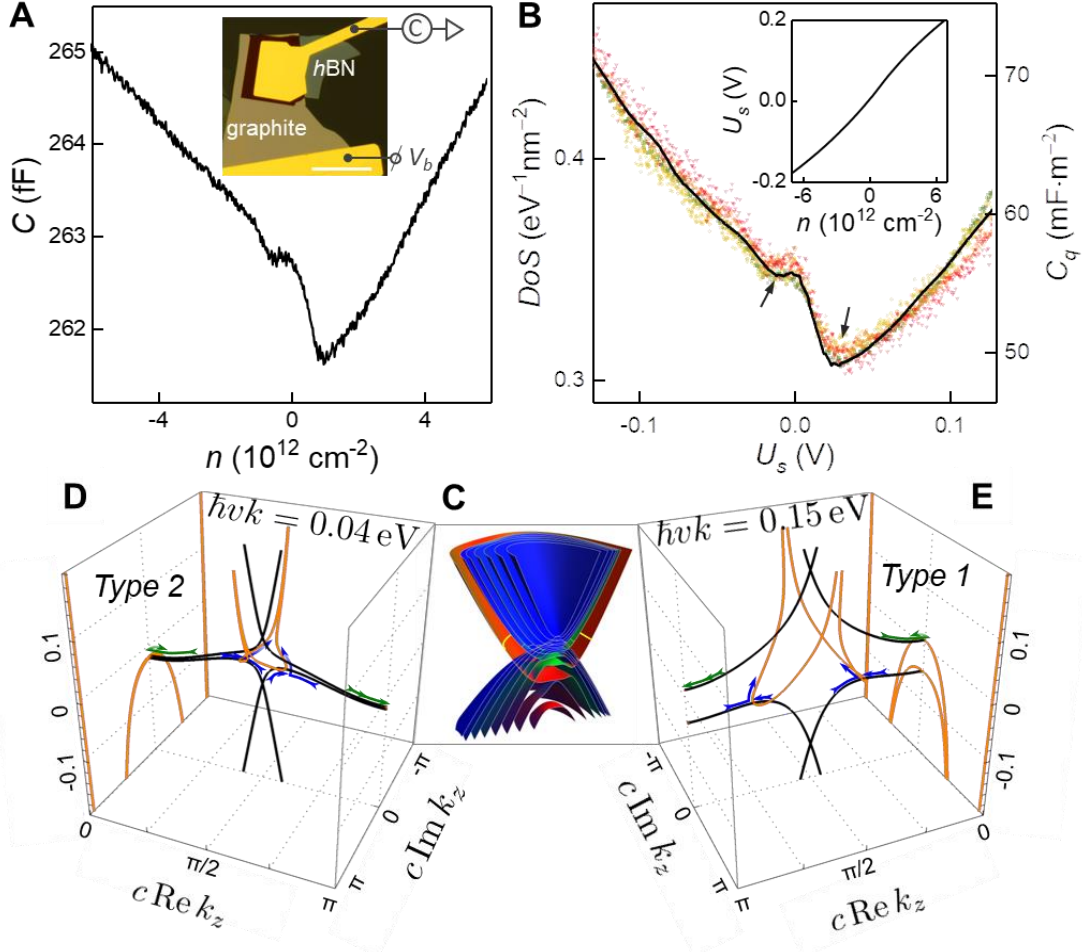

**Fig. S1.** *DoS* **spectroscopy of surface states of graphite in zero magnetic field.** (**A**) Capacitance (*C*) *vs* carrier density (*n*) at *T* = 0.3 K. Inset shows optical micrograph of one capacitor device (device D5), scale bar 20 μm. (**B**) Density of states (*DoS*) and quantum capacitance ($C_q$) *vs* surface potential ($U_s$) from calculations based on effective-mass model (black line). Coloured symbols are experimental data from four different devices. Inset is calculated $U_s$ *vs n*. (**C**) Calculated dispersion of a 14-layer-thick graphite film for electron doping $n = 6 \cdot 10^{12}$ cm$^{-2}$ as a function of in-plane momentum $k_x, k_y$ (horizontal axes) and energy *E* (vertical axis). Red (green) colour indicates surface states having high probability density of the wave function at the first (second) graphene bilayer. Upper outer surfaces with larger radius correspond to Type 1, and lower inner surface with a smaller radius correspond to Type 2. Blue colour indicates bulk states, and yellow contour highlights the Fermi level. (**D, E**) Dispersion for propagating (Im $k_z$ = 0, black lines) and evanescent modes (Im $k_z \neq 0$, orange lines) for bulk graphite as a function of complex $k_z$ for fixed $\hbar v k$ = 0.04 eV (panel (**D**), 1D metal regime with Type 2 evanescent modes) and 0.15 eV (panel (**E**), 1D semiconductor regime with Type 1 evanescent modes), respectively. Fermi level is at 0 eV. Green/blue arrows indicate evolution of surface states from propagating modes into evanescent modes for electron/hole doping.

To provide qualitative understanding of the surface states in graphite, we plot the eigenstates for homogenous bulk graphite in Fig. S1D, E, which consist of propagating (black curves, real $k_z$) and evanescent (orange curves, complex $k_z$) modes. There are three types of propagating modes: majority electron and hole bands with bandwidth $2\gamma_2$ and a minority carrier band near $ck_z = \pi/2$, $\gamma_2$ is the hopping amplitude between two non-dimer sites in the next-nearest layers. These propagating bands cross the bulk Fermi level for small in-plane momentum $k_x, k_y$ (Fig. S1D, 1D metal regime in z-direction), but are spread away from the Fermi level at large $k_x, k_y$ (Fig. S1E, 1D semiconductor regime in z-direction). When a potential near the surface is introduced by doping, the propagating modes in 1D semiconductor region start to cross the Fermi-level. With potential abating away from the surface, these modes evolve into evanescent modes in the gap, as shown by red/blue arrows in Fig. S1E for hole/electron doping, respectively. The dispersion of these evanescent modes, which we denote as Type 1, crosses the Fermi level, forming a surface Fermi-line with radius larger than Fermi surface of propagating carriers, as illustrated by the yellow contour in Fig. S1C. These states are similar to surface states in doped semiconductor, with the difference that they exist only for in-plane momenta larger than the projection of bulk Fermi surface of graphite. In 1D metal regime, another type of evanescent mode, denoted as Type 2, appears for $|E| > |\gamma_2|$ and never crosses the Fermi level, Fig. S1D.

Returning to the $C_q$-$U_s$ curve in Fig. S1B, at low doping when $|U_s| < |\gamma_2|$ (that is, $|n| < 6 \cdot 10^{11}$ cm$^{-2}$), there are no surface states, and quantum capacitance is determined by electron and hole screening. Since holes have a slightly larger *DoS* (shallower dispersion) than electrons, we see larger $C_Q$ at hole doping. When doping reaches $U_s \approx \pm\gamma_2$, Type 1 and Type 2 surface states appear and contribute to quantum capacitance. The radius of surface Fermi-line for Type 1 states, grows with $|n|$, leading to increasing density of surface states and growth of quantum capacitance.

## Supplementary Note 2 – Surface and bulk states in graphitic films in the presence of moiré superlattice

At low fields ($B < 1$ T), the onset of Shubnikov de Haas oscillations is strongly altered by the kaleidoscopic band structure of the surface states (Fig. 1E). In Fig. S2 we compare the low field transport for aligned and non-aligned devices, D1 and D4, of similar thickness (≈8 nm). In non-aligned graphite (Fig. S2 B,D), we observe that a Landau fan develops for finite densities $|n_b| > 0.6 \times 10^{12}$ cm$^{-2}$, and all QHE states can be traced back to $n_b \approx 0$ as $B$ approaches 0. In contrast, for aligned graphite (Fig. S2 A,C) similar QHE features are also overlaid by oscillations emanating from Lifshitz transitions at $|n| \approx 2$ and $|n| \approx 3.7 \times 10^{12}$ cm$^{-2}$ resulting in diamond-like features in $\sigma_{xx}$ occurring at flux fractions $\phi/\phi_0 = p/q$.

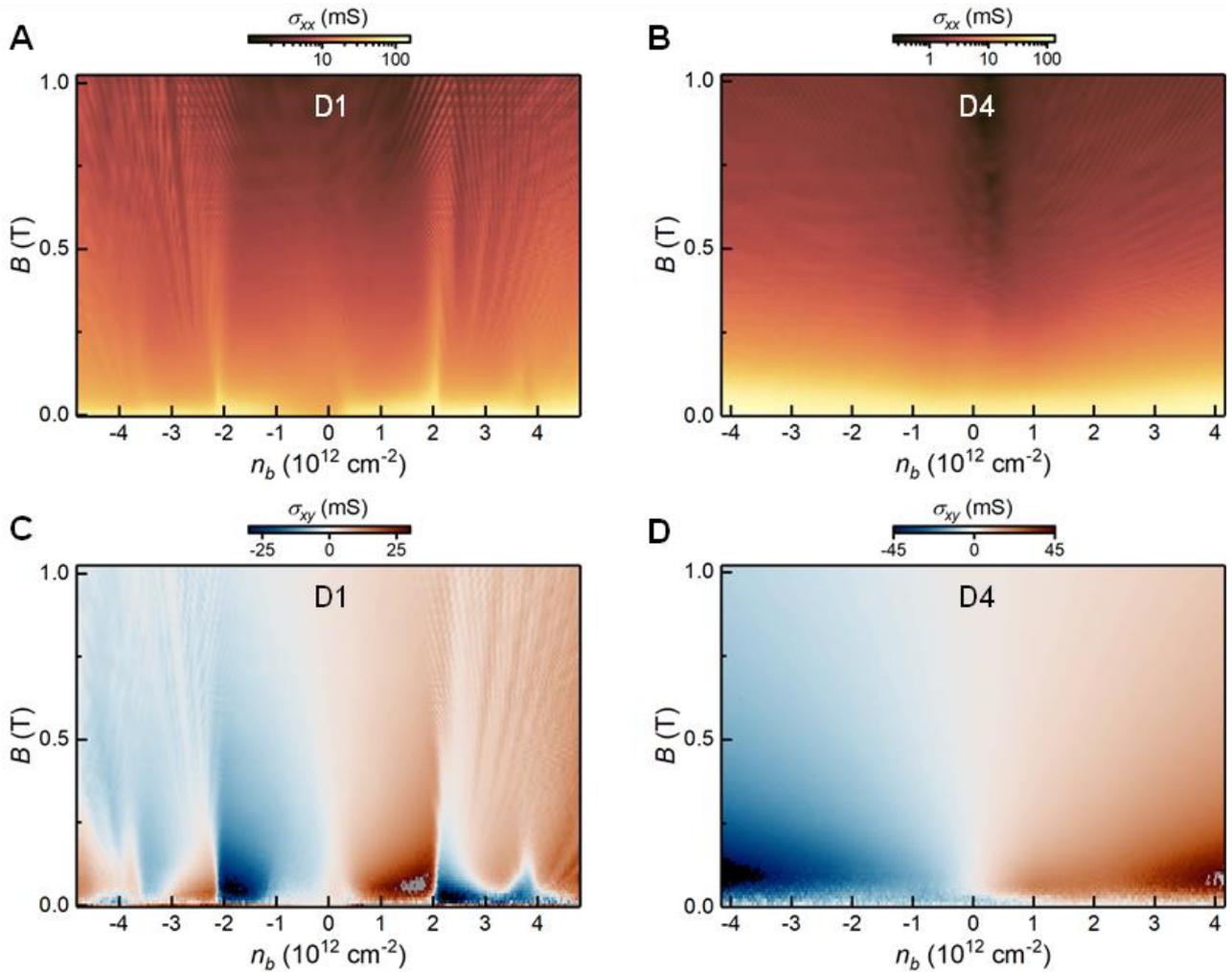

**Fig. S2. Transport in aligned vs non-aligned devices at small magnetic fields.** (**A, C**) $\sigma_{xx}$, $\sigma_{xy}$ as a function of $B$ and $n_b$ for aligned device D1 at $T = 0.24$ K. Landau level features emanate from LTs ($|n| \approx 2$ and $|n| \approx 3.7 \times 10^{12}$ cm$^{-2}$) and from zero doping. (**B, D**) $\sigma_{xx}$, $\sigma_{xy}$ as a function of $B$ and $n_b$ for non-aligned device D4 at $T = 0.22$ K.

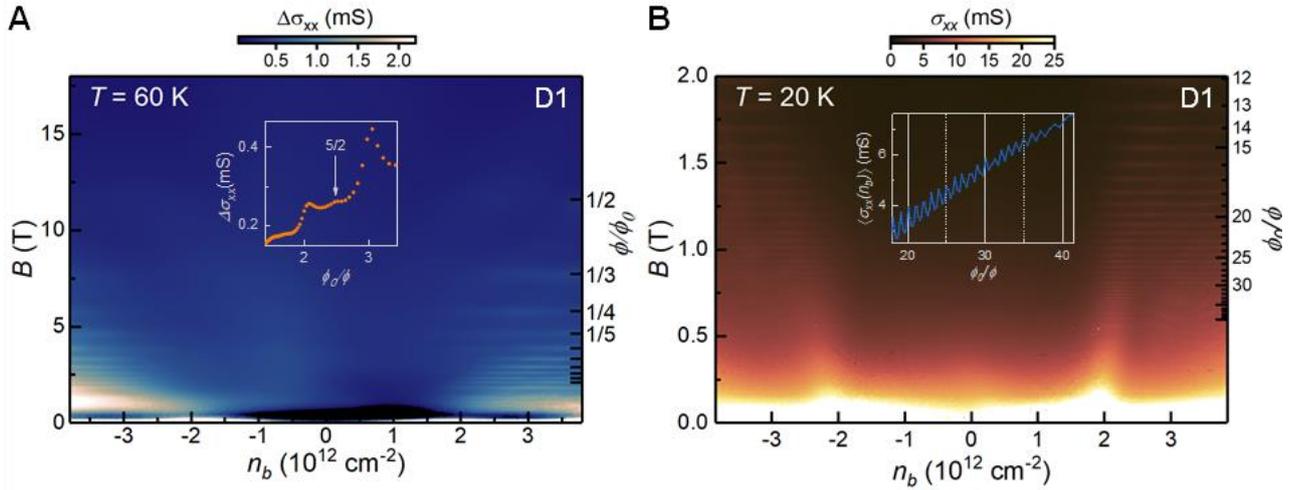

**Fig. S3. Extended range Brown-Zak mapping for device D1.** (**A**) Extended range plot of Fig. 2B; $\Delta\sigma_{xx}$ (conductivity minus a smooth background) as a function of $B$ and $n_b$ at T = 60 K. Inset, $\Delta\sigma_{xx}(\phi_0/\phi)$ trace at $n_b$ = 3.53 x$10^{12}$ cm$^{-2}$ highlighting higher order Brown-Zak oscillation at $\phi_0/\phi$ = 5/2. (**B**) Extended range plot of Fig. 2C; $\sigma_{xx}$ as a function of $B$ and $n_b$ at $T$ = 20 K. Inset, conductivity averaged across a carrier density range $n_b$ = 2.00 to 3.84 x$10^{12}$ cm$^{-2}$, ‹$\sigma_{xx}(n_b)$›, plotted as a function of $\phi_0/\phi$ showing oscillations continuing down to experimental mapping resolution ($B_{step}$ = 1 mT).

The methods of energy spectrum modelling applied to device D3 in the main text (Fig. 4) were also applied to device D2, which has a different alignment to hBN and layer parity. To be more specific, device D2 is 21-layer in thickness and only aligned to one encapsulating hBN. Layer parity plays a key role in QHE in graphite. The QHE exists as a result of standing waves in $k_z$ Landau bands, and the maxima of these standing waves is in the +KH valley for even layers and -KH valley for odd layers [26]. For even total number of layers ($N$ = 2$\mathcal{N}$, i.e., device D3), there are equal numbers of odd and even layers to host the states in +KH and -KH valleys, states therefore have a twofold degeneracy ($\Delta v$ = 2). However, in odd layer parity graphite films ($N$ = 2$\mathcal{N}$-1, i.e., device D2), the mismatch between number of even and odd layers lifts this degeneracy and the energy gaps between states are halved.

As shown in Fig. S4, device D2 exhibits main QHE sequence gaps (Fig. S4A-D) that are significantly reduced compared to that in D3 (Fig. 3E), due to the lifted ±KH valley degeneracy (Fig. S4E) because of the odd-layer parity of graphite in D2. In Fig. S4D, we focus on the zeroth gap size measured as a function of magnetic field between two level crossings at $B$ = 10 T and $B$ = 16 T, with a maximal observed gap of ≈0.48 meV. Fig. S4F plots the energy levels as a function of $B$ for 21-layer graphite, calculated in the same way as in Ref. [26]. Without moiré perturbation, both the extent of the zeroth gap (8.5 T < $B$ < 17 T) and its maximal size (1.3 meV) in the model are notably larger than in the experiment (cf. similar maximal gap size ≈1.1 meV calculated for 21 layers in Ref. [26]). Upon application of Hofstadter butterfly to each energy level (see methods) using the same scaling factor $S$ = 0.14 meV as in Fig. 3E in the main text, the zeroth gap in the model is reduced to ≈0.6 meV (Fig. S4G), in closer agreement with our experiments. However, such effective broadening of energy levels from the Hofstadter butterfly will lead to many overlapping states and hence gap closures, which were not observed in experiment. We postulate that this is due to inadequate treatment of equal moiré perturbation to states in both even and odd layers (+KH and -KH valleys). As only one of the $\mathcal{N}$ odd layers is directly affected by the Moiré potential, which rapidly decreases in strength with gaining depth into graphite, the perturbation strength on even layers is expected to be significantly reduced. In Fig. S4H, we plot a revised model with $S$ = 0.14 meV for odd layers and $S$ = 0.04 meV for even layers, yielding less main QHE gap closures than Fig. S4G while the same zeroth gap remains. Qualitatively, this is in a closer agreement with the experiment. Future work to further improve this model may use a more detailed treatment to Landau level mixing.

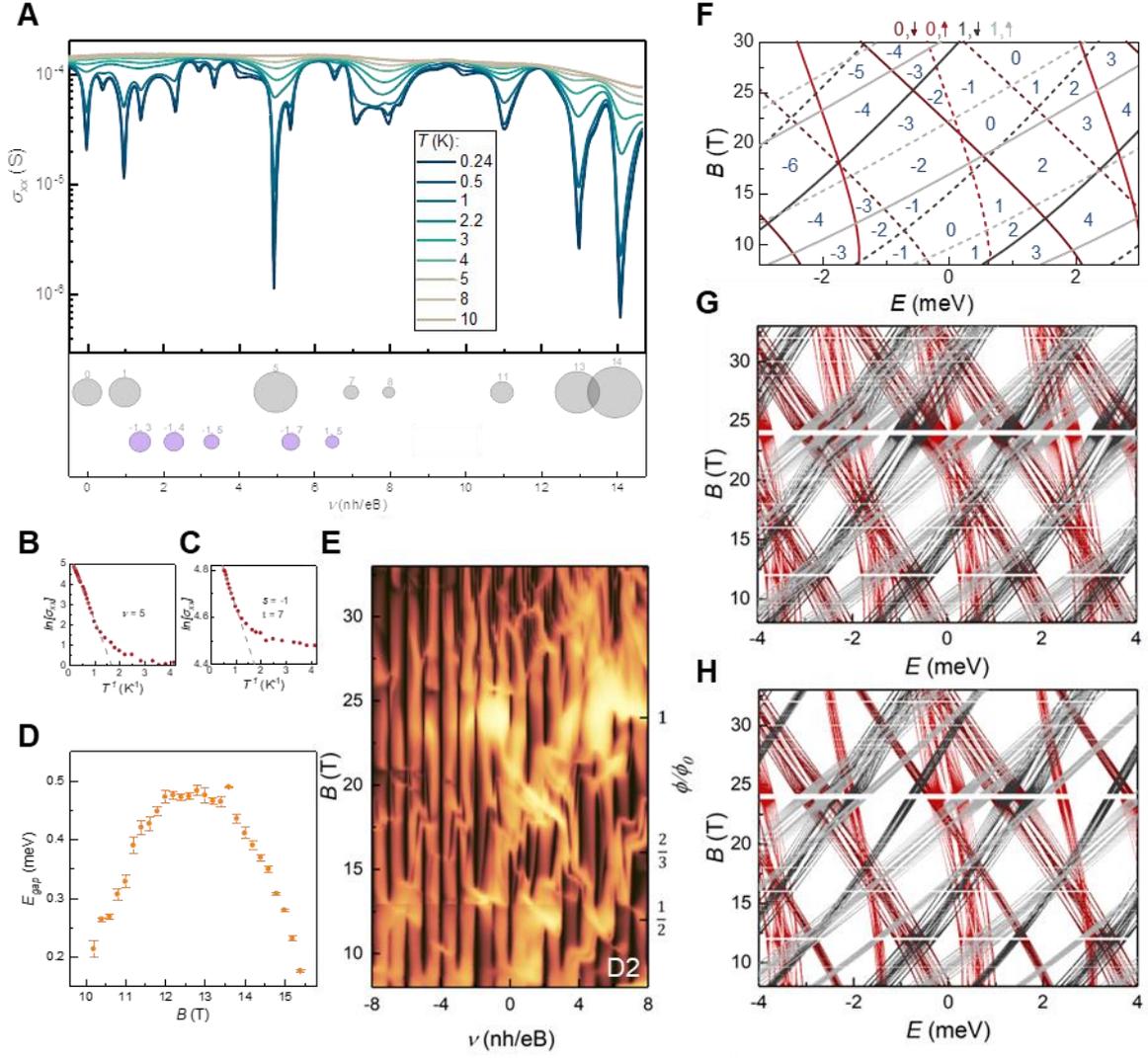

**Fig. S4. Gap size hierarchy in device D2.** (**A**) Upper panel, $\sigma_{xx}$ as a function of integer QHE filling factor $v$ at various temperatures, $B$ = 15.2 T. Lower panel, bubble plot of extracted gap size from Arrhenius fits to the $\sigma_{xx}$ minima. Gap size scales with area (60 μeV to 0.9 meV), and grey bubbles are integer gaps (labelled by $v$) and purple bubbles are fractal gaps (labelled by $s, t$). (**B-C**) Examples of Arrhenius plots of ln[$\sigma_{xx}$] as a function of reciprocal temperature for integer QHE gap at $v$ =5 and fractal QHE gap of integers ($s,t$)=(-1,7), respectively (these two states are highlighted by red circles in panel **A**). The linear regions are fitted to yield a slope of ~½$E_{gap}$. (**D**) Gap size extracted from Arrhenius fits as a function of $B$ for the zeroth gap. (**E**) Conductivity map for device D2, same data as in Fig. 3A in the main text, except plotted as a function of filling factor v of main QHE sequence. Colour scale brown to yellow is 0 to 210 μS. (**F**) Allowed energy levels resulting from quantised states from 0 (1) Landau bands shown in red (grey), calculated for 21-layer-thick graphite film without a moiré perturbation [26]. Zeeman splitting has been included, as indicated by light and dark lines for spin up and spin down, respectively. (**G**) Combination of panels (F) and Fig. 4A by applying the Hofstadter butterfly as a small perturbation ($S$ = 0.14 meV) to each energy level, Eq. 2. (**H**) Same as (G) but with $S$ = 0.14 meV for odd layer states and $S$ = 0.04 meV for even layer states.

# Supplementary Note 3 – Surface states in non-aligned graphite films in finite B-fields

In the presence of magnetic field perpendicular to the basal plane of graphite, surface states associated with graphite bulk Landau bands (BLBs) manifest in the capacitance spectra as pronounced magnetocapacitance oscillations, as shown in Fig. S5A.

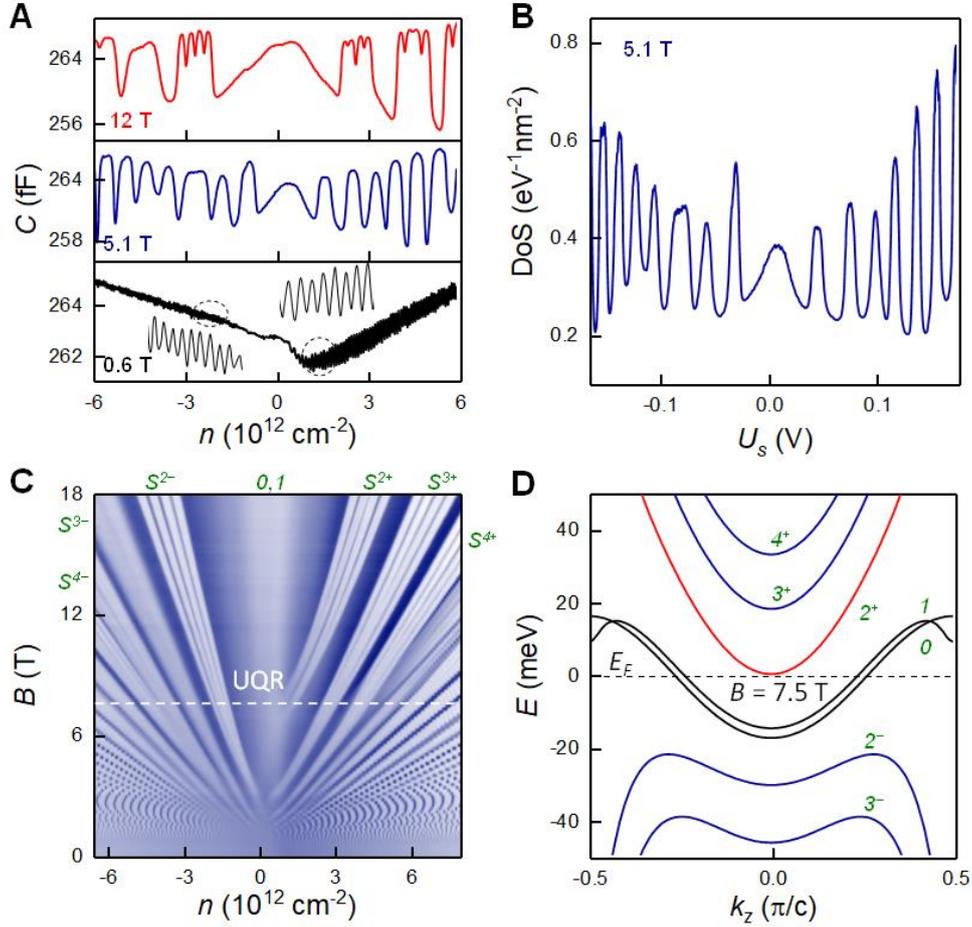

**Fig. S5. Magnetocapacitance oscillations and Landau fan diagram of the surface states.** (**A**) Typical $C(n)$ at 0.6 T (bottom black curve), 5.1 T (middle blue curve) and 12 T (top red curve), respectively, in capacitor device D6 at $T$ = 0.3 K. Bottom insets are magnifications of the marked areas. (**B**) $DoS$ vs $U_s$ at 5.1 T, 0.3 K. (**C**) Surface Landau fan diagram $C(n, B)$ at 0.3 K. Color scale: navy-to-white, 254.7 fF to 270.0 fF. Dashed line marks the critical field, where BLB $2^+$ leaves the Fermi level (d) and graphite enters the ultraquantum region (UQR). (**D**) Dispersion relation for BLBs calculated using the SWMC model at $B$ = 7.5 T.

For BLBs which still cross the Fermi level, the associated surface states would coexist and mix with the bulk states. However, BLBs of higher/lower energy can also become occupied at the surface by electrostatically doping, giving rise to surface Landau levels (SLLs). While filling these SLLs, regions of high compressibility appear as peaks in the capacitance spectra. Note that the width of these high-compressibility regions does not correspond to integer degeneracy (> 4), because some fraction of gate voltage induced charge is sunk into the bulk to support the self-consistent screening potential near the surface (Fig. S6).

Having determined the geometric capacitance from the fitting of zero field data, we can convert $C(n)$ into $DoS(U_s)$ via $U_s = eV_b - e^2n/C_G$ [41]. As shown in Fig. S5B, peaks in the $DoS$ correspond to metallic SLLs, which are separated by relatively low $DoS$ regions (cyclotron gaps of the surface states). In contrast to classical 2D systems, the $DoS$ in these cyclotron gaps is non-zero, because charges can be injected into the bulk graphite. At 12 T, three minima are further developed on top of most peaks, indicating that the spin and valley degeneracy of the SLLs is lifted.

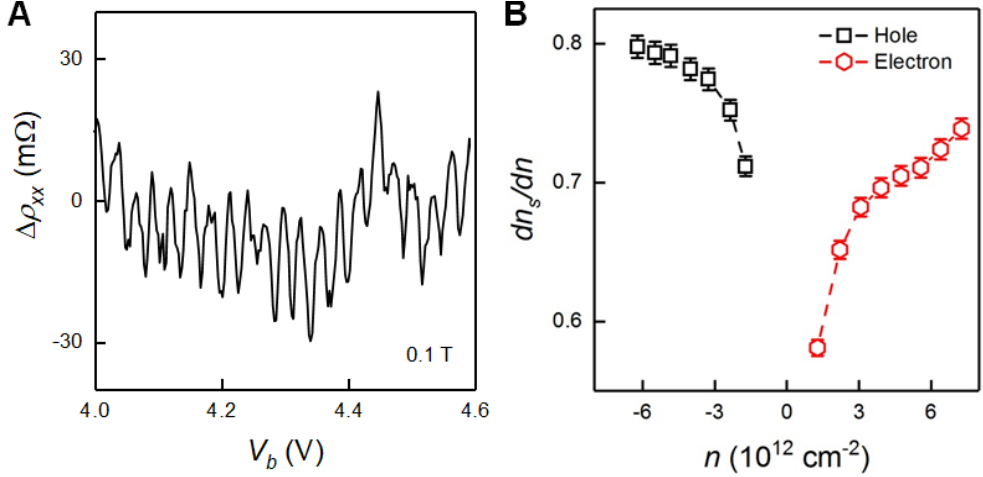

**Fig. S6. Low field quantum oscillations.** (**A**) Oscillations of $\Delta\rho_{xx}$ in device D7 obtained by subtracting a smooth background while sweeping the gate voltage ($V_b$) at 0.1 T, 0.3 K using excitation $I$ = 1 µA. The encapsulated graphite with a thickness of 20 nm is defined to Hall bar geometry for this measurement. (**B**) $dn_s/dn$ as a function of $n$ at 0.6T, where $n_s$ is the carrier density injected into the surface states, $n$ is the total electrostatically induced carrier density. It is deduced from the curve shown in Fig. S5A, based on the periodicity of the oscillations.

Experimental results are better visualized and more informative when presented as $C(n,B)$ map, Fig. S5C. Note the branches of surface states spawn out from the neutrality point at $B$ = 7.5 T, 3 T, 2 T, and so on. These $B$-fields correspond to the critical fields above which the BLBs no longer cross the Fermi level and appear only as SLLs. For instance, according to our SWMC model, at 7.5 T, BLB $2^+$ is just above the Fermi level (Fig. S5D). Thus, a branch of surface states spawned out around this field is labelled as $S^{2+}$. The same happens with the electron BLB $3^+$ at 3 T and hole BLB $2^-$ at 2T, Fig. S5C. Note that above 7.5 T, only BLB 0 and 1 cross the Fermi level at charge neutrality point and graphite enters the ultraquantum regime (UQR) [48, 49]. The broad feature in the middle region of the map, whose position does not change notably with $B$ is also attributed to the BLB 0, 1 [50, 51].

We observed oscillations down to $B \approx 0.1$ T, see Fig. S5C and Fig. S6A, which sets a lower bound for surface charge carrier mobility of $\approx$ 100,000 cm$^2$V$^{-1}$s$^{-1}$. The high electronic quality of surface states also enables fractional features in Landau quantization of charge carriers. Another graphite capacitor device was therefore fabricated to investigate this, with a thicker hBN dielectric to reduce the inhomogeneity of electrostatic potential from the metal electrode. At high magnetic field, $B$ = 20 T, we observe the formation of two minima on top of singly degenerate states of $S^{2+}$ (Fig. S7A). We plotted d$C$/d$v$ as a function of filling factor $v$ at high field region (Fig. S7B). The $\Delta v$ between the fractional gap is around 0.27, which is lower than expected $\Delta v$ = 1/3 for fractional QHE.

To further characterize these fractional states, a graphite Hall bar device with a 6 nm thick graphite was fabricated, with graphite not aligned with hBN. In this thin graphite device, the formation of standing waves leads to bulk QHE gaps in the UQR [26]. Through applying a displacement field, $B$-$n$ regions can be found where the energy level of surface states locates within the bulk gap (Fig. S8). In these regions, the surface states are isolated from the bulk completely, and both vanishing longitudinal conductivity ($\sigma_{xx}$) and quantized Hall conductivity ($\sigma_{xy}$) clearly indicate the development of fractional QHE with a 1/3 degeneracy. The difference between the capacitance and transport measurements can be reconciled, when considering the negative compressibility of the fractional states: chemical potential of the surface states reduces with the injection of $n$ [52-54], acquiring additional charges from the bulk. Not only $S^{2+}$, but also $S^{2-}$ shows a similar behaviour (Fig. S9), indicating a strong electron-electron interaction in the surface states.

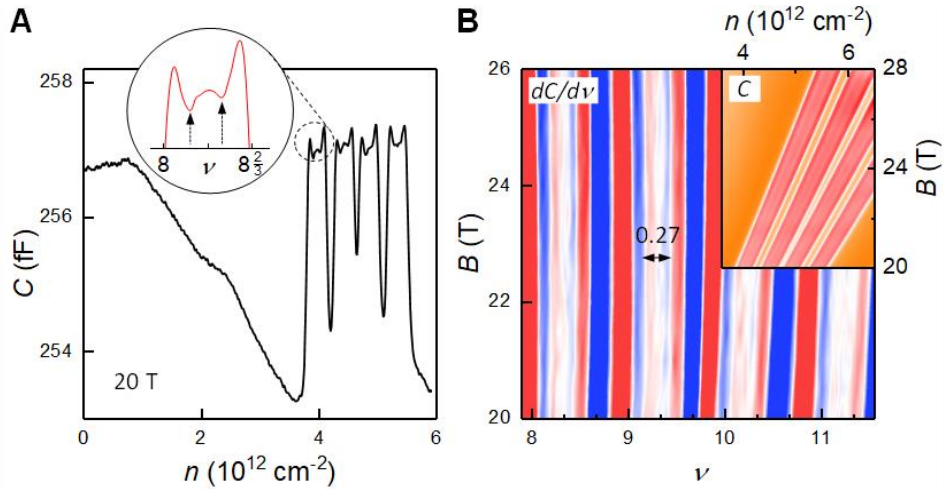

**Fig. S7. Fractional surface states in capacitance measurements.** (**A**) $C(n)$ at 20 T, $T$ = 0.3 K measured in capacitor device D8. Inset magnifies the circled region, but plots as $C(\nu)$. (**B**) $dC/d\nu(\nu, B)$ of $S^{2+}$ at high field region. Colour scale: blue-white-red, -6 fF to 6 fF. Right-top inset shows the corresponding $C$ ($n, B$) map. Colour scale: orange-to-red, 254.3 fF to 260 fF.

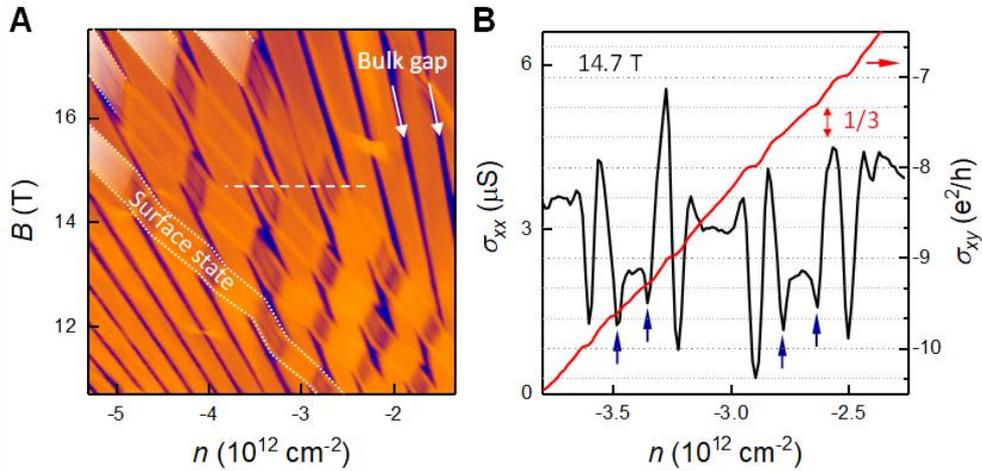

**Fig. S8. Fractional surface states in transport measurements.** (**A**) Longitudinal conductivity $\sigma_{xx}$ (B, n) measured at $D = (n_t-n_b)\cdot e/2\varepsilon_0$ = 0.24 V/nm in a Hall bar device D9 (9-nm-thick-graphite), $T$ = 0.3 K, $I$ = 20 nA. The white shaded areas are guides to the eye for the surface states. Boundaries of one such surface states are marked by white dotted curves. Logarithmic colour scale, navy to orange is 0.1 to 118.2 μS. (**B**) $\sigma_{xx}$ cut profile (black curve) of the white dashed line in (A) and the corresponding Hall conductivity $\sigma_{xy}$ (red curve).

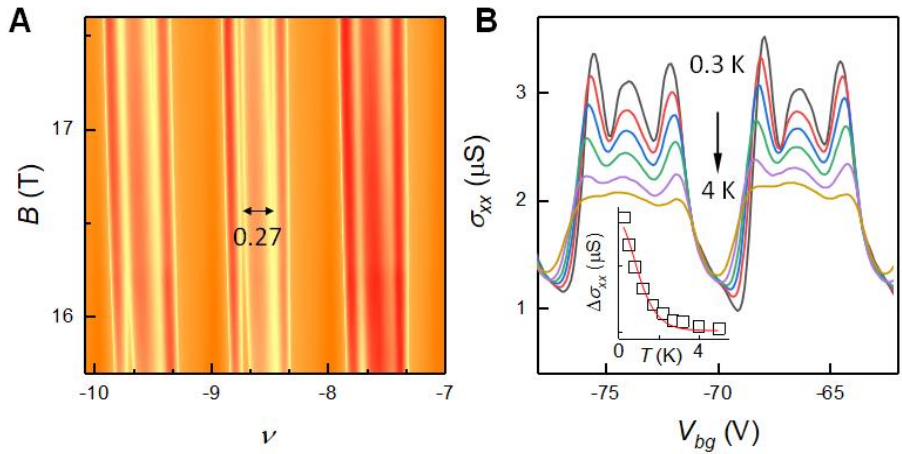

**Fig. S9. Fractional states of $S^{2-}$.** (**A**) $\sigma_{xx}$ ($\nu, B$) map of a 5-7 nm graphite in Corbino geometry (device D10), $T$ = 0.3 K. Color scale: orange-to-red, 0 to 1 μS. (**B**) $\sigma_{xx}(V_{bg})$ versus temperature at 18 T. Inset is $\Delta\sigma_{xx}$ ($T$) and its fitting based on Lifshitz-Kosevich model.

## Supplementary Note 4 – Raman spectroscopy of aligned graphite films

To characterize the effect of surface superlattice potential on hBN-encapsulated graphite we performed Raman spectroscopy measurements. A graphite flake with extended monolayer graphene (MLG) region was selected to benchmark the alignment of the entire graphite film, Fig. S10A,B. Raman spectra of MLG/hBN superlattice has been well studied [55], and the alignment can be traced by the width of a 2D peak of MLG. The 2D peak of MLG broadens with better alignment, due to the increased strain inhomogeneity caused by moiré periodic potential from hBN substrate. Similar broadening of 2D peak was also observed in bilayer graphene (BLG)/hBN superlattice system [56], indicating that superlattice potential from hBN substrate is able to propagate through graphene bilayers, thus detectable by Raman spectroscopy. However, how far such superlattice potential can penetrate into the bulk of graphite remains elusive.

To clarify this, we fabricated two hBN/Grt/hBN heterostructures at the same time, by transferring graphite onto two adjacent but intentionally misoriented hBN flakes. Graphite flake is controlled to be aligned with one of the hBNs, consequently, is misaligned to the other, as shown in Fig. S10C. The flake alignment is characterized by the full width at half maximum (FWHM) of MLG 2D peak, Fig. S10E. Each spectrum was averaged over ten spectra acquired at different positions and normalized by the intensity of $E_{2g}$ hBN peak at 1363 cm$^{-1}$. The FWHM is 21 cm$^{-1}$ and 35 cm$^{-1}$ for non-aligned and aligned regions of MLG, respectively, which agrees well with the results in Ref. [55]. Broadening of the 2D peak is expected if the superlattice potential at the interface can propagate through the bulk graphite crystal. However, we found no appreciable difference on the Raman map of 2D FWHM between aligned and non-aligned graphite regions, Fig. S10D,E. This implies that the surface superlattice potential from hBN substrate is not able to penetrate through our graphite flake, which is ≈2.6 nm thick. Since the thickness of graphite flakes used in main text is no less than 5 nm, it is therefore, safe to conclude that the graphite flakes are not likely to be affected by the periodic potential of hBN at the interface.

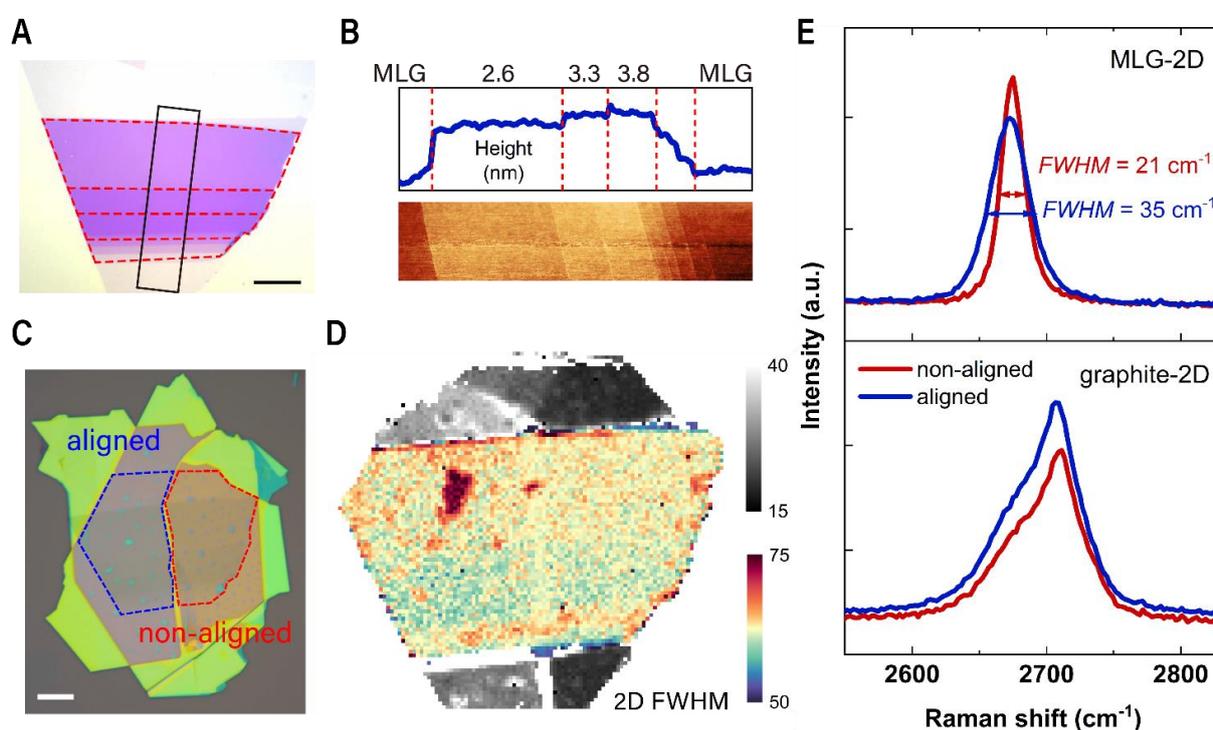

**Fig. S10. Raman characterization of aligned and non-aligned ABA graphite.** (**A**) Optical image and (**B**) AFM profile of graphite flake used in Raman measurement. The flake contains both monolayer graphene (MLG) region and graphite region with thickness around 10 layers. Scale bar 10 um. (**C**) Optical image of the stack fabricated for Raman measurement. The aligned and non-aligned region are marked by blue and red dashed lines. Scale bar 10 um. (**D**) Raman map of full width half maximum (FWHM) of 2D peak. The MLG and graphite region are color-coded in grayscale and red-purple respectively. (**E**) Comparison of 2D peak between aligned and non-aligned regions in MLG (up) and graphite (down). Each spectrum shown here is averaged over ten spectra at different spots. The intensity is normalized by the intensity of hBN peak at 1363 cm$^{-1}$.